\documentclass[longauth]{aa}
\usepackage{graphicx}
\usepackage{txfonts}
\usepackage{xcolor}
\usepackage{hyperref}
\hypersetup{colorlinks, linkcolor=blue, citecolor=blue, urlcolor=blue}
\newcommand{\muvec}{\mbox{\boldmath $\mu$}}

\newcommand{\te}{t_{\rm E}}
\newcommand{\thetae}{\theta_{\rm E}}

\newcommand{\pie}{\pi_{\rm E}}

\newcommand{\dl}{D_{\rm L}}
\newcommand{\ds}{D_{\rm S}}

\definecolor{brown}{rgb}{0.59, 0.29, 0.0}
\definecolor{darkgreen}{rgb}{0.0, 0.42, 0.24}
\definecolor{darkblue}{rgb}{0.01, 0.31, 0.59}
\definecolor{darkblue}{rgb}{0.0, 0.25, 0.42}
\definecolor{blue}{rgb}{0.0,0.0,1.0}
\definecolor{green}{rgb}{0.0,1.0,0.0}

\begin{document}

\title{MOA-2022-BLG-563Lb, KMT-2023-BLG-0469Lb, and KMT-2023-BLG-0735Lb: Three
sub-Jovian-mass microlensing planets}
\titlerunning{MOA-2022-BLG-563Lb, KMT-2023-BLG-0469Lb, and KMT-2023-BLG-0735Lb}

\author{
% leading author -----------------------------
     Cheongho~Han\inst{\ref{01}}
\and Youn~Kil~Jung\inst{\ref{02},\ref{03}}
\and Ian~A.~Bond\inst{\ref{04}}
\and Andrew~Gould\inst{\ref{05},\ref{06}}
\\
(Leading authors)\\  
% KMTNet ---------------------------
     Michael~D.~Albrow\inst{\ref{07}}
\and Sun-Ju~Chung\inst{\ref{02},\ref{08}}
\and Kyu-Ha~Hwang\inst{\ref{02}}
\and Chung-Uk~Lee\inst{\ref{02}}
\and Yoon-Hyun~Ryu\inst{\ref{02}}
\and In-Gu~Shin\inst{\ref{08}}
\and Yossi~Shvartzvald\inst{\ref{09}}
\and Hongjing~Yang\inst{\ref{10}}
\and Jennifer~C.~Yee\inst{\ref{08}}
\and Weicheng~Zang\inst{\ref{08},\ref{10}}
\and Sang-Mok~Cha\inst{\ref{02},\ref{11}}
\and Doeon~Kim\inst{\ref{01}}
\and Dong-Jin~Kim\inst{\ref{02}}
\and Seung-Lee~Kim\inst{\ref{02}}
\and Dong-Joo~Lee\inst{\ref{02}}
\and Yongseok~Lee\inst{\ref{02},\ref{11}}
\and Byeong-Gon~Park\inst{\ref{02}}
\and Richard~W.~Pogge\inst{\ref{06}}
\\
(The KMTNet collaboration)\\
%% MOA ---------------------------
     Fumio~Abe\inst{\ref{12}}
\and Richard~Barry\inst{\ref{13}}
\and David~P.~Bennett\inst{\ref{13},\ref{14}}
\and Aparna~Bhattacharya\inst{\ref{13},\ref{14}}
\and Hirosame~Fujii\inst{\ref{12}}
\and Akihiko~Fukui\inst{\ref{12},\ref{16}}
\and Ryusei~Hamada\inst{\ref{17}}
\and Yuki~Hirao\inst{\ref{17}}
\and Stela~Ishitani Silva\inst{\ref{13},\ref{18}}
\and Yoshitaka~Itow\inst{\ref{12}}
\and Rintaro~Kirikawa\inst{\ref{17}}
\and Naoki~Koshimoto\inst{\ref{17}}
\and Yutaka~Matsubara\inst{\ref{12}}
\and Shota~Miyazaki\inst{\ref{19}}
\and Yasushi~Muraki\inst{\ref{12}}
\and Greg~Olmschenk\inst{\ref{13}}
\and Cl{\'e}ment~Ranc\inst{\ref{20}}
\and Nicholas~J.~Rattenbury\inst{\ref{21}}
\and Yuki~Satoh\inst{\ref{17}}
\and Takahiro~Sumi\inst{\ref{17}}
\and Daisuke~Suzuki\inst{\ref{17}}
\and Mio~Tomoyoshi\inst{\ref{17}}
\and Paul~J.~Tristram\inst{\ref{22}}
\and Aikaterini~Vandorou\inst{\ref{13},\ref{14}}
\and Hibiki~Yama\inst{\ref{17}}
\and Kansuke~Yamashita\inst{\ref{17}}
\\
(The MOA Collaboration)\\
}

\institute{
      Department of Physics, Chungbuk National University, Cheongju 28644, Republic of Korea \label{01}                                                            % (01)
\and  Korea Astronomy and Space Science Institute, Daejon 34055, Republic of Korea \label{02}                                                                      % (02)
\and  Korea University of Science and Technology, Korea, (UST), 217 Gajeong-ro, Yuseong-gu, Daejeon, 34113, Republic of Korea \label{03}                           % (03)
\and  Institute of Natural and Mathematical Science, Massey University, Auckland 0745, New Zealand \label{04}                                                      % (04)
\and  Max-Planck-Institute for Astronomy, K\"{o}nigstuhl 17, 69117 Heidelberg, Germany \label{05}                                                                  % (05)
\and  Department of Astronomy, Ohio State University, 140 W. 18th Ave., Columbus, OH 43210, USA \label{06}                                                         % (06)
\and  University of Canterbury, Department of Physics and Astronomy, Private Bag 4800, Christchurch 8020, New Zealand \label{07}                                   % (07)
\and  Center for Astrophysics $|$ Harvard \& Smithsonian, 60 Garden St., Cambridge, MA 02138, USA \label{08}                                                       % (08)
\and  Department of Particle Physics and Astrophysics, Weizmann Institute of Science, Rehovot 76100, Israel \label{09}                                             % (09)
\and  Department of Astronomy, Tsinghua University, Beijing 100084, China \label{10}                                                                              % (10)
\and  School of Space Research, Kyung Hee University, Yongin, Kyeonggi 17104, Republic of Korea \label{11}                                                        % (11)
\and  Institute for Space-Earth Environmental Research, Nagoya University, Nagoya 464-8601, Japan \label{12}                                                      % (12)
\and  Code 667, NASA Goddard Space Flight Center, Greenbelt, MD 20771, USA \label{13}                                                                             % (13)
\and  Department of Astronomy, University of Maryland, College Park, MD 20742, USA  \label{14}                                                                    % (14)
\and  Department of Earth and Planetary Science, Graduate School of Science, The University of Tokyo, 7-3-1 Hongo, Bunkyo-ku, Tokyo 113-0033, Japan\label{15}     % (15)
\and  Instituto de Astrof{\'i}sica de Canarias, V{\'i}a L{\'a}ctea s/n, E-38205 La Laguna, Tenerife, Spain  \label{16}                                            % (16)
\and  Department of Earth and Space Science, Graduate School of Science, Osaka University, Toyonaka, Osaka 560-0043, Japan \label{17}                             % (17)
\and  Department of Physics, The Catholic University of America, Washington, DC 20064, USA \label{18}                                                             % (18)
\and  Institute of Space and Astronautical Science, Japan Aerospace Exploration Agency, 3-1-1 Yoshinodai, Chuo, Sagamihara, Kanagawa 252-5210, Japan\label{19}    % (19)
\and  Sorbonne Universit\'e, CNRS, UMR 7095, Institut d'Astrophysique de Paris, 98 bis bd Arago, 75014 Paris, France \label{20}                                   % (20)
\and  Department of Physics, University of Auckland, Private Bag 92019, Auckland, New Zealand \label{21}                                                          % (21)
\and  University of Canterbury Mt.~John Observatory, P.O. Box 56, Lake Tekapo 8770, New Zealand \label{22}                                                        % (22)
}

%\date{Received ; accepted}

% \abstract{}{}{}{}{} 
% 5 {} token are mandatory
\abstract
{}
{
We analyze the anomalies appearing in the light curves of the three microlensing events
MOA-2022-BLG-563, KMT-2023-BLG-0469, and KMT-2023-BLG-0735. The anomalies exhibit common 
short-term dip features that appear near the peak.
}
{
From the detailed analyses of the light curves, we find that the anomalies were produced by
planets accompanied by the lenses of the events. For all three events, the estimated mass ratios 
between the planet and host are on the order of $10^{-4}$: $q\sim 8 \times 10^{-4}$ for MOA-2022-BLG-563L, 
$q\sim 2.5\times 10^{-4}$ for KMT-2023-BLG-0469L, and $q\sim 1.9\times 10^{-4}$ for KMT-2023-BLG-0735L. 
The interpretations of the anomalies are subject to a common inner-outer degeneracy, which causes 
ambiguity when estimating the projected planet-host separation. 
}
{
We estimated the planet mass, $M_{\rm p}$, host mass, $M_{\rm h}$, and distance, $D_{\rm L}$, 
to the planetary system by conducting Bayesian analyses using the observables of the events. 
The estimated physical parameters of the planetary systems are 
$(M_{\rm h}/M_\odot, M_{\rm p}/M_{\rm J}, D_{\rm L}/{\rm kpc}) = 
(0.48^{+0.36}_{-0.30}, 0.40^{+0.31}_{-0.25}, 6.53^{+1.12}_{-1.57})$ for MOA-2022-BLG-563L,
$(0.47^{+0.35}_{-0.26}, 0.124^{+0.092}_{-0.067}, 7.07^{+1.03}_{-1.19})$ for KMT-2023-BLG-0469L, and
$(0.62^{+0.34}_{-0.35}, 0.125^{+0.068}_{-0.070}, 6.26^{+1.27}_{-1.67})$ for KMT-2023-BLG-0735L.
According to the estimated parameters, all planets are cold planets with projected separations that 
are greater than the snow lines of the planetary systems, they have masses that lie between the masses 
of Uranus and Jupiter of the Solar System, and the hosts of the planets are main-sequence stars that 
are less massive than the Sun.  In all cases, the planetary systems are more likely to be in the bulge 
with probabilities $P_{\rm bulge} = 64\%$, 73\%, and 56\% for MOA-2022-BLG-563, KMT-2023-BLG-0469, 
and KMT-2023-BLG-0735, respectively. 
}
{}


\maketitle

\section{Introduction} \label{sec:one}

Because of its trait that does not depend on the luminosity of a lensing object, microlensing
provides an important tool to detect extrasolar planets. Planetary microlensing \citep{Mao1991, 
Gould1992} was proposed before the first discoveries of an exoplanet belonging to a pulsar
\citep{Wolszczan1992} and a planet belonging to a normal stellar system \citep{Mayor1995}, 
but the first microlensing planet was only discovered in 2004 \citep{Bond2004}, that is, 12 
years after the operation of the first-generation microlensing surveys, for example, Massive 
Compact Halo Objects \citep[MACHO:][]{Alcock1993}, Optical Gravitational Lensing Experiment 
\citep[OGLE:][]{Udalski1994}, and Exp\'erience pour la Recherche d'Objets Sombres 
\citep[EROS:][]{Aubourg1993}.  The difficulty of finding planets from these early microlensing 
surveys was mostly ascribed to the low cadence of the surveys, which were typically carried out 
with a 1~day cadence, while the duration of a planet-induced lensing signal is several hours 
for terrestrial planets and several days even for giant planets. In order to increase the 
observational cadence, planetary microlensing experiments during the time from the mid 1990s to 
mid 2010s had been carried out in a hybrid mode \citep{Gould1992, Albrow1998}, in which survey 
experiments with a low observational cadence mainly focused on detecting lensing events, and 
follow-up groups conducted high-cadence observations for a fraction of events detected by the 
survey experiments.

% Table 1 ------------------------------------------------
\begin{table*}[t]
\small
%\centering
\caption{Coordinates, baseline magnitude, and extinction.  \label{table:one}}
\begin{tabular}{lllll}
%\begin{tabular}{\columnwidth}{@{\extracolsep{\fill}}lllcc}
\hline\hline
\multicolumn{1}{c}{Event}                      &
\multicolumn{1}{c}{(RA, DEC)$_{\rm J2000}$}    &
\multicolumn{1}{c}{$(l,b)$}                    &
\multicolumn{1}{c}{$I_{\rm base}$}             &
\multicolumn{1}{l}{$A_I$}                     \\
\hline
 MOA-2022-BLG-563  & (18:01:12.05, -27:50:08.99) & $(2^\circ\hskip-2pt .687, -2^\circ\hskip-2pt .395)$  &  18.12 &  1.25  \\
 KMT-2023-BLG-0469 & (17:57:37.63, -28:42:54.22) & $(1^\circ\hskip-2pt .533, -2^\circ\hskip-2pt .149)$  &  18.45 &  1.50  \\
 KMT-2023-BLG-0735 & (17:50:18.23, -29:12:10.51) & $(0^\circ\hskip-2pt .301, -1^\circ\hskip-2pt .014)$  &  19.79 &  5.97  \\
\hline
\end{tabular}
%\tablefoot{ ${\rm HJD}^\prime = {\rm HJD}- 2450000$.  }
\end{table*}
% --------------------------------------------------------

Planetary microlensing experiments entered a new phase in the mid 2010s by dramatically increasing 
the observational cadence of lensing surveys with the employment of globally distributed multiple 
telescopes that are equipped with very wide-field cameras.  With the upgraded instrument, the 
current lensing surveys achieve observational cadences down to 0.25 hr, which is two orders of 
magnitude shorter than the cadence of the early experiments. The operation of the high-cadence 
surveys has led to the great increase in the event detection rate, and the number of lensing events 
that are annually detected by the current lensing surveys is more than 3000, which is nearly two 
orders of magnitude higher than the rate of the early surveys. Being able to monitor a large number 
of lensing events with enhanced cadences, the detection rate of planets has also greatly increased. 
According to the NASA Exoplanet Archive\footnote{\tt exoplanetarchive.ipac.caltech.edu}, the number 
of microlensing planets that have been detected since the full operation of the high-cadence surveys 
in 2016 is 137, and this is more than two times the 64 planets found from the previous experiments 
that had been conducted for 24 years from the year 1992 to 2015.  Considering that the high-cadence 
surveys were partially operated in 2020 due to the COVID pandemic and the data collected in the 2022 
and 2023 season are under analysis, the current surveys are annually detecting about 30 planets on 
average \citep{Gould2022, Jung2022}.

The growing number of microlensing planets reveals similar anomaly patterns in certain planetary 
signals. \citet{Han2023} outlined instances of planetary signals emerging via a non-caustic-crossing 
channel in their analysis of two lensing events KMT-2022-BLG-0475 and KMT-2022-BLG-1480. Similarly, 
\citet{Han2021a} depicted analogous occurrences of non-caustic-crossing signals in their study of 
the three planetary lensing events KMT-2018-BLG-1976, KMT-2018-BLG-1996, and OGLE-2019-BLG-0954. In 
their examination of the the three planetary events KMT-2017-BLG-2509, OGLE-2017-BLG-1099, and 
OGLE-2019-BLG-0299, \citet{Han2021c} presented various instances where planetary signals emerged 
from source crossings over resonant caustics formed by giant planets near the Einstein rings of 
host stars. \citet{Jung2021} demonstrated planetary signals arising from perturbations by peripheral 
caustics induced by planets in their analyses of two lensing events OGLE-2018-BLG-0567 and 
OGLE-2018-BLG-0962.  Moreover, in the cases of KMT-2018-BLG-1743 and KMT-2021-BLG-1898, \citet{Han2021b} 
and \citet{Han2022}, respectively, highlighted two microlensing planets with signals deformed by companions 
to the source stars. Additionally, \citet{Han2017} presented an exemplary case of a planetary signal 
arising through a repeating channel from the analysis of the  microlensing planet OGLE-2016-BLG-0263Lb. 
These various studies collectively underscore the diverse mechanisms and complexities observed in 
detecting microlensing planets.

Microlensing planets are found from detailed analyses of anomalous signals appearing in numerous 
lensing light curves. These analyses are done via a complex procedure and require heavy computations 
in order to identify the planetary origin of signals by distinguishing them from signals of other 
origins. Therefore, morphological studies by classifying lensing events with similar anomalies and 
investigating the origins for the individual classes of anomalies are important not only for the 
diagnosis of anomalies before detailed analysis but also for the accurate characterization of 
lens systems for future lensing events with similar anomaly structures.

In this work, we conduct analyses of the three compiled planetary microlensing events that were
detected from short-term signals exhibiting similar anomaly features, including MOA-2022-BLG-563
KMT-2023-BLG-0469 and KMT-2023-BLG-0735. These events share similar characteristics, in which the 
anomalies showed up near the peaks of the lensing light curves and they exhibit similar dip 
features surrounded by peaks on both sides of the dips. We present the common characteristics of 
the planetary systems that are found from the detailed analyses of the anomalies.

We present the analyses of the lensing events according to the following organization.  In 
Sect.~\ref{sec:two}, we describe the observations conducted to acquire the data used in the analyses, 
and mention the instruments used for the observations and the procedure of data reduction and error-bar 
normalization.  In Sect.~\ref{sec:three}, we begin by briefly mentioning planetary lensing properties and 
introducing parameters used in the modeling of the lensing light curves.  In the subsequent subsections, 
we present detailed analyses conducted for the individual events: in Sect.~\ref{sec:three-one} for 
MOA-2022-BLG-563, in Sect.~\ref{sec:three-two}   for KMT-2023-BLG-0469, and in Sect.~\ref{sec:three-three} 
for KMT-2023-BLG-0735.  In Sect.~\ref{sec:four}, we characterize the source stars of the events and 
estimate the angular Einstein radii.  In Sect.~\ref{sec:five}, we determine the physical parameters of 
the planetary system by conducting Bayesian analyses for the individual events.  We summarize the results 
found from the analyses and conclude in Sect.~\ref{sec:six}.

\section{Observations and data}\label{sec:two}

The anomalous nature of the three lensing events MOA-2022-BLG-563, KMT-2023-BLG-0469, and
KMT-2023-BLG-0735 were found from the inspection of the data  obtained from the two high-cadence
lensing surveys conducted toward the Galactic bulge field by the Korea Microlensing Telescope Network 
\citep[KMTNet:][]{Kim2016} group and the Microlensing Observations in Astrophysics \citep[MOA:][]{Bond2001} 
group. Among these events, KMT-2023-BLG-0469 and KMT-2023-BLG-0735 were observed solely by the KMTNet 
group, and MOA-2022-BLG-563 was observed by both survey groups.  The KMTNet ID reference corresponding 
to the event MOA-2022-BLG-563 is KMT-2022-BLG-2681, and, hereafter, we  use the MOA ID reference because 
the event was first discovered by the MOA group. In Table~\ref{table:one}, we list the equatorial and 
galactic coordinates of the events together with the $I$-band baseline magnitudes, $I_{\rm base}$, and 
extinction, $A_I$, toward the source stars.  The $I$-band extinction is estimated as  $A_I = 7\,A_K$, 
where the $K$-band extinction $A_K$ is adopted from \citet{Gonzalez2012}.

The KMTNet observations of the events were carried out using three identical 1.6 m telescopes, 
which are globally distributed in the three continents of the Southern Hemisphere: at the Cerro
Tololo Inter-American Observatory in Chile, the South African Astronomical Observatory in South
Africa, and the Siding Spring Observatory in Australia. We designate the individual KMTNet
telescopes as KMTC, KMTS, and KMTA using the initials of the countries in which the telescopes
are located. The MOA observations were done using the 1.8~m telescope of the Mt. John Observatory 
located in New Zealand. The fields of view of the cameras mounted on the KMTNet telescopes and the 
MOA telescope are 4~deg$^2$ and 2.2~deg$^2$, respectively. We describe details of the observations 
conducted for the individual events in the following section.

Reduction of images and photometry of the source stars were carried out using the photometry codes 
of the individual groups, which were developed by \citet{Albrow2009} for the KMTNet survey and by 
\citet{Bond2001} for the MOA survey. For a fraction of the KMTC data set, additional photometry 
using the pyDIA code \citep{Albrow2017} was done to construct color-magnitude diagram (CMD) of 
stars lying near the source stars and to determine the source colors. For the data used in the 
analyses, the error bars estimated from the photometry pipelines were readjusted so that they 
are consistent with the scatter of the data and $\chi^2$ per degree of freedom (dof) for each 
data set becomes unity. This error-bar normalization process was done following the routine depicted 
in detail by \citet{Yee2012}.

\section{Anomaly analysis}\label{sec:three}

The image of a microlensed source star is split into two, in which the brighter one (major image)
appears outside of the Einstein ring, and the fainter one (minor image) appears inside the ring.
Planet-induced anomalies in a lensing light curve arise when a planet lies close to one of these
two source images. The "major-image perturbation" refers to the case in which the planet perturbs
the major image, and the "minor-image perturbation" indicates the case in which the planet
perturbs the minor image \citep{Gaudi1997}. When the planet perturbs the major image, the
image is further magnified by the planet, and this results in a bump feature (positive deviations)
of the anomaly. When the planet disturbs the minor image, by contrast, the image is demagnified,
and this results in a dip feature (negative deviations) of the anomaly. Because the major image
lies outside the Einstein ring, the bump feature of a planetary anomaly is generally produced by a 
"wide planet", for which the projected planet-host separation is greater than the Einstein radius. 
By contrast, the dip feature of an anomaly is generated by a "close planet", for which the separation 
is smaller than the Einstein radius, because the minor image lies inside the Einstein ring. See the
appendix of \citet{Han2018} for more detailed discussion on the types of planetary perturbations.

As viewed on the source plane, planet-induced anomalies arise when the source approaches the
planet-induced caustics. Caustics refer to the source positions at which the lensing magnifications
of a point source are infinite. Planet-induced caustics form two sets of caustics, in which the
"central caustic" forms near the position of the planet host, and the "planetary caustic" forms 
away from the host at the position $\sim (1 - 1/s^2){\bf s}$.  Here ${\bf s}$ denotes the position 
vector of the planet from the host with a length normalized to the angular Einstein radius $\thetae$ 
of the planetary lens system.  For detailed descriptions on the position, size, and shape of the 
central and planetary caustics, see \citet{Chung2005} and \citet{Han2006}.

There exist multiple known channels that can produce short-term anomalies in lensing light curves. 
Besides the planetary channel, it is known that bump-featured anomalies can be produced via a 1L2S 
channel, in which the lens system is composed of a single-mass lens and  binary source, and the 
bump feature can be generated by the close approach of the faint source companion to the lens 
\citep{Gaudi1998}. Dip-featured anomalies can be produced not only by a planet but also by a binary 
companion to the lens, although the two features can be usually distinguished from the shape of the
 dip feature \citep{Han2008}.

The events analyzed in this work share a common characteristics that short-term anomalies with
dip features appear near the peaks of the lensing light curves.  Considering that a dip-featured 
anomaly is produced by a lens with either a binary companion or a planet, we conduct a binary-lens 
single-source (2L1S) modeling of the light curves for the interpretations of the observed anomalies. 
For each lensing event, the modeling was done in search of a lensing solution, which refers to a 
set of the lensing parameters that define the light curve of the event. Under the approximation of 
the rectilinear relative motion between the lens and source, the light curve of a 2L1S event is 
characterized by 7 lensing parameters. The first set of the three parameters $(t_0, u_0, \te)$ 
define the source approach to the lens, and the individual parameters denote the time of the closest 
lens-source approach, the separation at that time (impact parameter), and the event time scale, 
respectively. The second set of the three parameters $(s, q, \alpha)$ define the binary lens, and 
the first two parameters represent the projected separation and mass ratio between the binary-lens 
components ($M_1$ and $M_2$), respectively, and the other parameter denotes the angle (source 
trajectory angle) between the direction of the relative lens-source proper motion vector $\muvec$ and 
the $M_1$--$M_2$ axis.  The lengths of $u_0$ and $s$ are scaled to the Einstein radius.  The last 
parameter $\rho$ is defined as the ratio of the angular source radius $\theta_*$ to the Einstein 
radius, that is, $\rho=\theta_*/\thetae$ (normalized source radius), and it describes the deformation 
of a lensing light curve by finite-source effects during source crossings over caustics or close 
approaches to caustics.

Among the lensing parameters, we search for the binary parameters $s$ and $q$ via a grid approach 
with multiple starting values of $\alpha$, and find the other parameters via a downhill approach.  
In the downhill approach, $\chi^2$ is minimized using the Markov Chain Monte Carlo (MCMC) method 
with an adaptive step size Gaussian sampler \citep{Doran2004}.  We then refine the local solutions 
appearing in the $\Delta\chi^2$ map on the plane of the grid parameters, and then find a global 
solution by comparing the $\chi^2$ values of the individual local solutions.  The anomalies of the 
all analyzed events appear near the peaks of the lensing light curves.  Such central anomalies are 
known to be produced by a planetary companion lying around the Einstein ring of the planet host or 
a very close or wide binary companion to the primary lens.  In order to check both the planetary 
and binary origins of the anomalies, we set the ranges of the grid wide enough to check both 
possible origins of the anomalies: $-1.0 < \log s \leq 1.0$ and $-5.0 < \log q \leq 1.0$.  For the 
case that multiple solutions with similar $\chi^2$ values exist, we present all degenerate solutions.

In the following subsections, we present the analyses conducted for the individual events. We
describe in detail the features of the anomalies residing in the light curves and present the 
results found from the modeling.  For events with multiple identified solutions, we explain the 
cause of the degeneracy among the solutions.

% Figure 1 ------------------------------------------------------
\begin{figure}[t]
\includegraphics[width=\columnwidth]{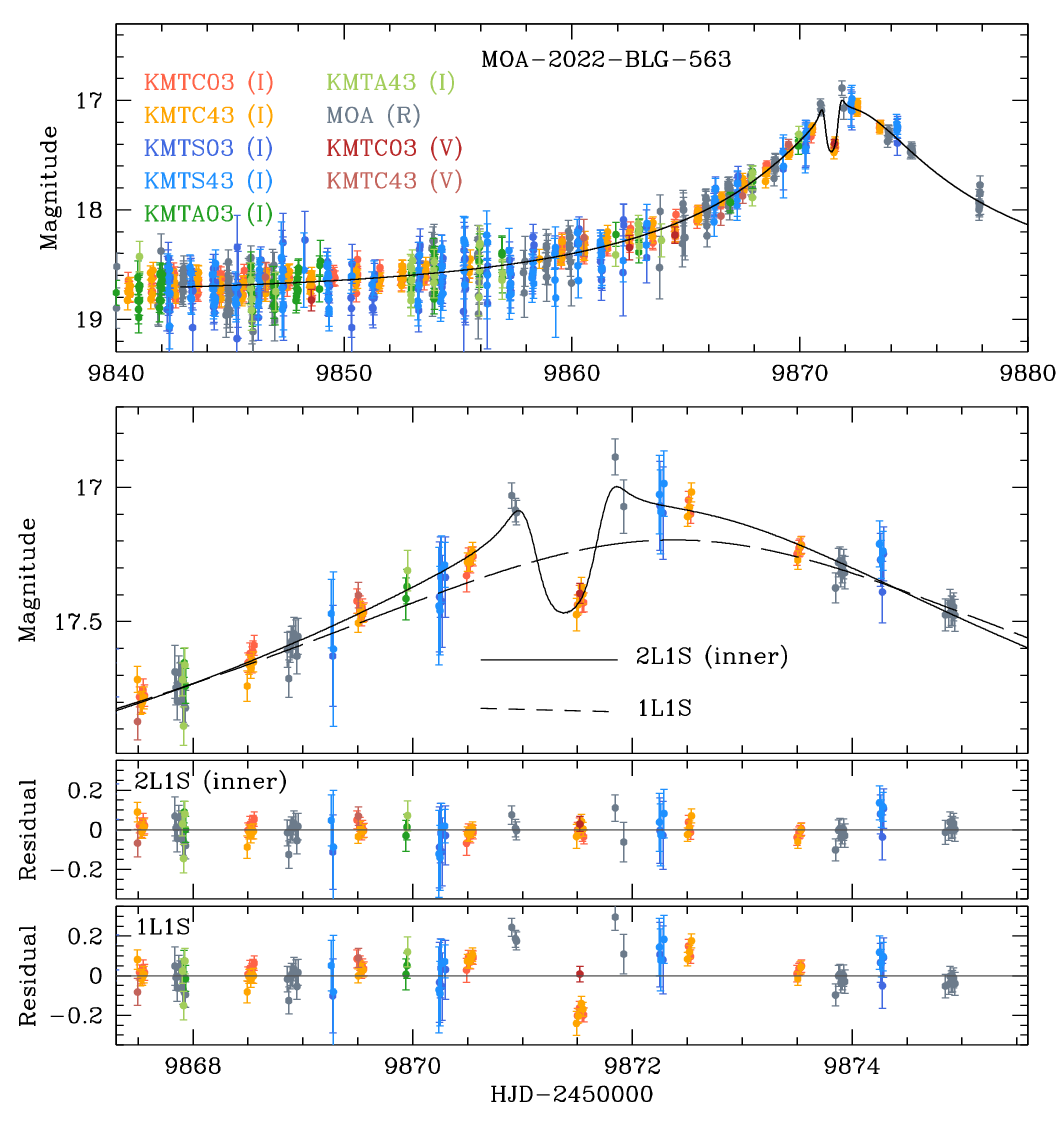}
\caption{
Light curve of MOA-2022-BLG-563. The top panel shows the whole view, and the lower panels show 
the zoom-in view around the anomaly and the residuals from the 2L1S and 1L1S models.  The solid 
and dashed curves drawn over the data points are the 2L1S and  1L1S models, respectively.  The 
colors of data points are set to match the labels of the observatories in the legend.
}
\label{fig:one}
\end{figure}
% --------------------------------------------------------------

\subsection{MOA-2022-BLG-563 }\label{sec:three-one}

The lensing event MOA-2022-BLG-563 was first found on 2022 October 14, which corresponds to 
the abridged Heliocentric Julian day ${\rm HJD}^\prime \equiv {\rm HJD}-2450000\sim 9866$,
by the MOA group, and was later identified by the KMTNet group from the post-season analysis of 
the 2022 season data \citep{Kim2018}.  The source position of the event corresponds to 
the KMTNet prime fields of BLG03 and BLG43, toward which observations were conducted with a 0.5~hr 
cadence for each field and a 0.25~hr cadence in combination. The MOA observations were conducted 
with a similar cadence.  The event occurred during the late stage of the 2022 bulge season, and 
thus the declining part of the light curve after ${\rm HJD}^\prime \sim 9878$ could not be covered.  
The source brightness returned to the baseline in the beginning of the 2023 season, and thus we do 
not include the 2023 season data in the analysis.  We included the KMTC $V$-band data in the analysis 
because one data point lies in the region of the anomaly. Although the event lies in the field 
that usually is observed frequently, there are gaps among the data sets because of the short 
duration of observation time at the late stage of the bulge season.

The lensing light curve of MOA-2022-BLG-563 is shown in Figure~\ref{fig:one}.  The event reached 
a moderate magnification of $A_{\rm max}\sim 5.5$ at the peak. From the inspection of the light 
curve, we find that the peak region of the light curve displays a short-term deviation from a 
single-lens single-source (1L1S) model. The anomaly is characterized by a dip feature centered at 
${\rm HJD}^\prime \sim 9871.5$ and double peaks lying on both sides of the dip. The time gap 
between the peaks is about $0.8$~day. The dip part of the anomaly was covered by the KMTC data, 
and the peaks were covered by the MOA data. Besides these main features, the data before and after 
the peaks show weak positive deviations from the 1L1S model.

% Figure 2 ------------------------------------------------------
\begin{figure}[t]
\includegraphics[width=\columnwidth]{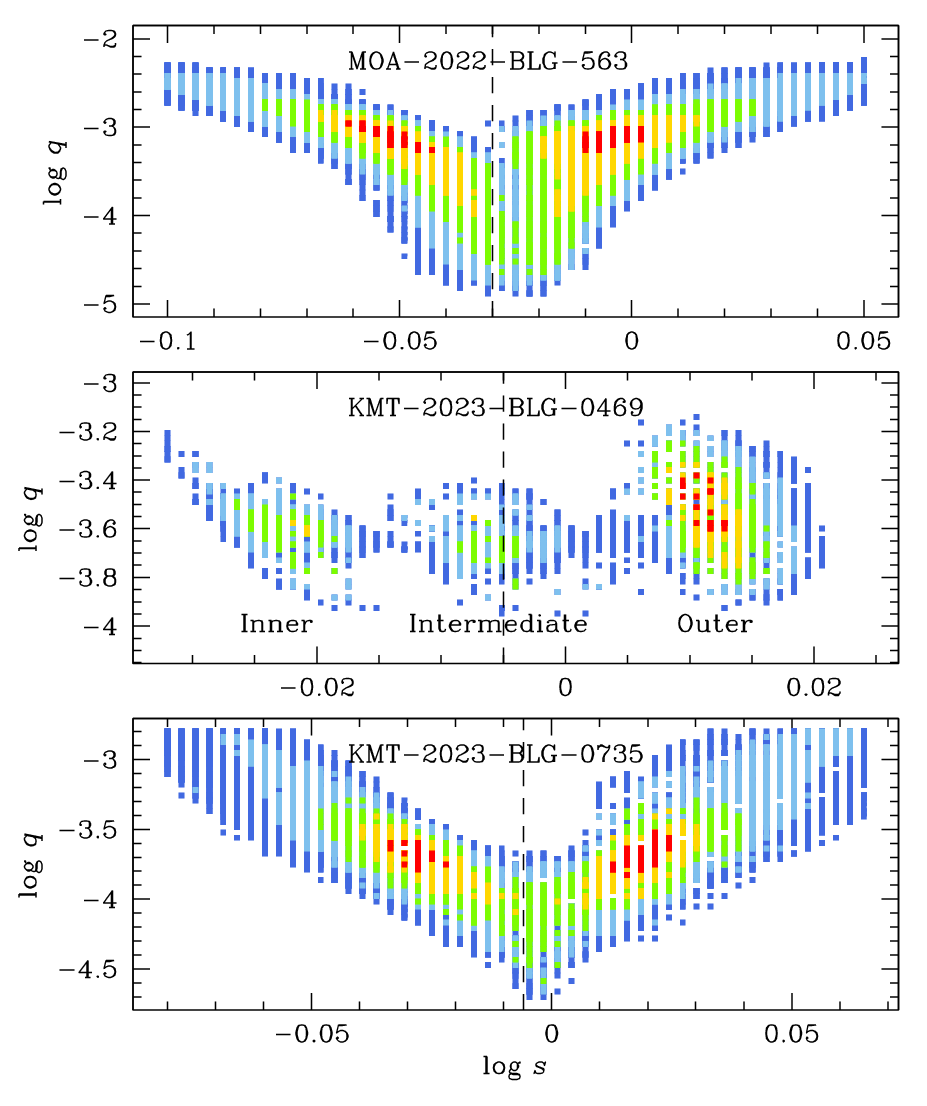}
\caption{
$\Delta\chi^2$ maps on the $(\log s, \log q)$ parameter plane for the events MOA-2022-BLG-563 
(top panel), KMT-2023-BLG-0469 (middle panel), and KMT-2023-BLG-0735 (bottom panel). Color 
coding is set to indicate points with 
$\Delta\chi^2 \leq 1^2n$ (red), 
$\leq 2^2n$ (yellow), 
$\leq 3^2n$ (green), 
$\leq 4^2n$ (cyan), and 
$\leq 5^2n$ (blue), where $n=2$. The dashed vertical line in each panel represents the geometric 
mean, $(s_{\rm in}\times s_{\rm out})^{1/2}$, of the planet separations of the inner and outer 
solutions. 
}
\label{fig:two}
\end{figure}
% --------------------------------------------------------------

% Figure 3 ------------------------------------------------------
\begin{figure}[t]
\includegraphics[width=\columnwidth]{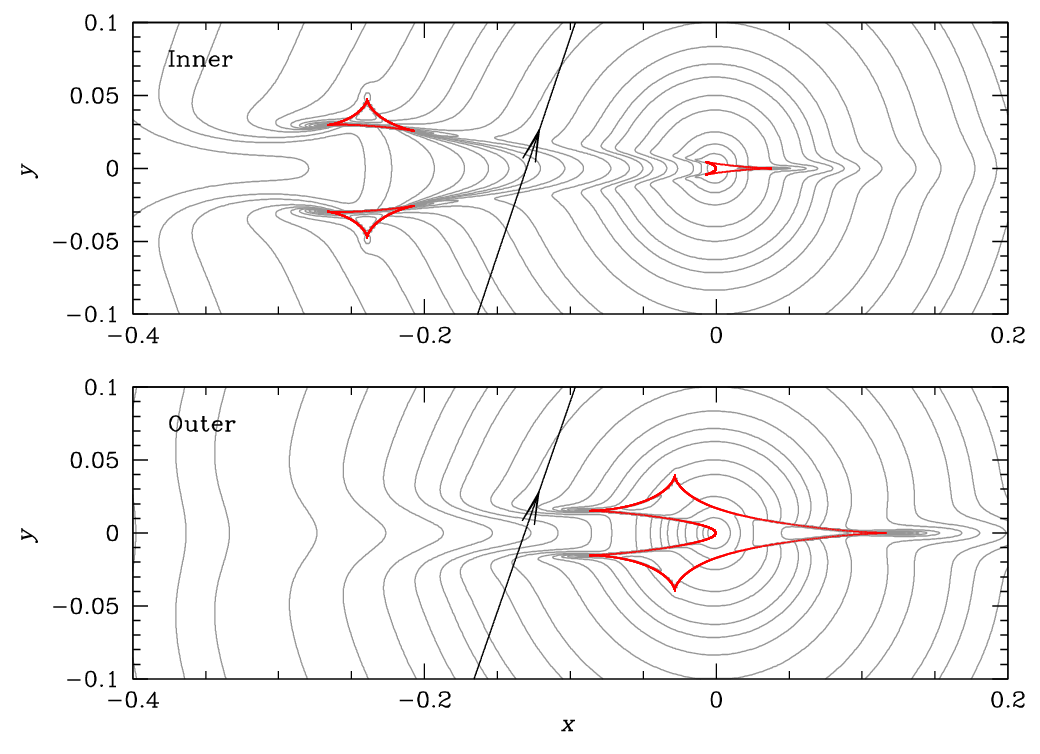}
\caption{
Lens-system configurations of MOA-2022-BLG-563.  The configurations of the inner and outer 
solutions are shown in the upper and lower panels, respectively.  In each panel, the red 
figures represent the caustics, and the arrowed line is the source trajectory. Gray contours 
encompassing the caustics represent equi-magnification contours.  The coordinates are centered 
at the center of mass of the planetary lens system, and the lengths are scaled to the Einstein 
radius.
}
\label{fig:three}
\end{figure}
% --------------------------------------------------------------

% Table 2 ------------------------------------------------
\begin{table}[t]
\small
%\centering
\caption{Model parameters of MOA-2022-BLG-563.\label{table:two}}
\begin{tabular*}{\columnwidth}{@{\extracolsep{\fill}}llll}
\hline\hline
\multicolumn{1}{c}{Parameter}    &
\multicolumn{1}{c}{Inner}        &
\multicolumn{1}{c}{Outer}       \\
\hline
$\chi^2$/dof            &   $ 5535.7/5355        $       &   $5536.7/5355        $      \\
$t_0$ (HJD$^\prime$)    &   $ 9872.226 \pm 0.041 $       &   $9872.214 \pm 0.040 $      \\
$u_0$                   &   $ 0.123 \pm 0.009    $       &   $0.124 \pm 0.009    $      \\
$\te$ (days)            &   $ 19.72 \pm 0.92     $       &   $19.58 \pm 0.97     $      \\
$s$                     &   $ 0.886 \pm 0.009    $       &   $0.986 \pm 0.010    $      \\
$q$ ($10^{-4}$)         &   $ 8.06 \pm 1.55      $       &   $7.41 \pm 1.79      $      \\
$\alpha$ (rad)          &   $ 1.896 \pm 0.022    $       &   $1.905 \pm 0.025    $      \\
$\rho$ ($10^{-3}$)      &   --                           &   --                         \\
\hline
\end{tabular*}
\tablefoot{ ${\rm HJD}^\prime = {\rm HJD}- 2450000$.  }
\end{table}
% --------------------------------------------------------

From the modeling of the light curve, it was found that the anomaly was produced by a planetary
companion to the lens. We identified a pair of degenerate solutions with binary parameters 
$(\log s, \log q)_{\rm in}\sim (-0.050, -3.1)$ and 
$(\log s, \log q)_{\rm out}\sim (-0.005, -3.1)$.
The top panel of Figure~\ref{fig:two} shows 
the positions of the two local solutions in the $\Delta\chi^2$ map on the $(\log s, \log q)$ 
parameter plane. For the reason discussed below, we designate the former and latter solutions 
as "inner" and "outer" solutions, respectively. In Table~\ref{table:two}, we list the full 
lensing parameters of the inner and outer solutions together with $\chi^2$ values of the fits 
and degrees of freedom (dof). The two solutions are very degenerate, and the inner solution is 
favored over the outer solution by only $\Delta\chi^2=1.0$. The model curve of the inner 
solution is drawn over the data points in Figure~\ref{fig:one}, and the residuals from the 
inner 2L1S and 1L1S solutions in the region around the anomaly are shown in the lower two 
panels.

The upper and lower panels of Figure~\ref{fig:three} show the lens-system configurations of the
inner and outer solutions, respectively. The configurations show that the source passed the region 
between the central and planetary caustics according to the inner solution, while the source passed 
the region outside the caustic according to the outer solution.  The planetary and central caustics 
of the inner solution are well separated, while the two sets of caustics of the outer solution merge 
together and form a single resonant caustic.  The dip feature of the anomaly was produced when 
the source passed through the negative deviation region on the back-end side of the caustic, and the 
bump features were produced when the source went through the positive deviation region extending from 
the strong cusps of the caustic.  The normalized source radius $\rho$ could have been measured  if 
the event had occurred in mid-season, so that there would be dense coverage over the cusp approach, 
but the large gaps among the data sets together with the non-caustic-crossing nature of the anomaly 
features made it difficult to measure $\rho$.

We find that the degeneracy between the two solutions originates from the "inner-outer" degeneracy. 
This degeneracy was originally introduced by \citet{Gaudi1997} to point out the similarity between 
the planetary signals resulting from the source passages through the near (inner) and far (outer) 
sides of the planetary caustic.  \citet{Yee2021} pointed out the continuous transition between the 
inner-outer and close-wide degeneracies, where the latter degeneracy originates from the similarity 
between the central caustics induced by a pair of planets with separation $s$ and $1/s$ 
\citep{Griest1998, Dominik1999, An2005}.  While the planetary separations of the pair of solutions 
that are subject to the close-wide degeneracy follows the relation $s_{\rm close} \times s_{\rm wide}
=1$, the planetary separations of the pair of solutions that are subject to the inner-outer degeneracy, 
$s_{\rm in}$ and $s_{\rm out}$, follow the relation
\begin{equation}
s_\pm^\dagger = (s_{\rm in}\times s_{\rm out})^{1/2} = 
{\sqrt{u_{\rm anom}^2+4} \pm u_{\rm anom}\over 2},
\label{eq1}
\end{equation}
where $u_{\rm anom} = (\tau _{\rm anom}^2+ u_0^2)^{1/2}$, $\tau_{\rm anom}=(t_{\rm anom}-t_0)/\te$, 
$t_{\rm anom}$ indicates the time of the anomaly, and the signs in the right term are "+" and "$-$" 
for the anomalies that are generated by the major and minor images perturbations, respectively 
\citep{Hwang2022, Gould2022}.  We note that for $u_{\rm anom}\rightarrow 0$, we have $s^\dagger_\pm 
\rightarrow 1$, which recovers the high-magnification limit of \citet{Griest1998}.  In the case 
of MOA-2022-BLG-563, the sign is "$-$" because the anomaly displayed a dip feature produced by 
the minor-image perturbation.  With the lensing parameters $(t_0, u_0, \te, t_{\rm anom}) \sim 
(9872.226, 0.123, 19.6, 9871.3)$, we find $s^\dagger \sim 0.936$, and this matches very well the 
geometric mean $(s_{\rm in}\times s_{\rm out})^{1/2} \sim 0.935$.  In the top panel of 
Figure~\ref{fig:two}, we mark the $s^\dagger$ as a dashed vertical line.

% Table 3 ------------------------------------------------
\begin{table*}[t]
\small
%\centering
\caption{Model parameters of KMT-2023-BLG-0469. \label{table:three}}
\begin{tabular}{lllll}
\hline\hline
\multicolumn{1}{c}{Parameter}      &
\multicolumn{1}{c}{Inner}          &
\multicolumn{1}{c}{Intermediate}   &
\multicolumn{1}{c}{Outer}         \\
\hline
$\chi^2$/dof            &   $6664.6/6668         $    &   $6676.6/6668          $     &  $6652.3/6668         $    \\
$t_0$ (HJD$^\prime$)    &   $10052.069 \pm 0.005 $    &   $10052.080 \pm 0.005  $     &  $10052.081 \pm 0.005 $    \\
$u_0$ ($10^{-3}$)       &   $8.42 \pm 0.20       $    &   $8.24 \pm 0.22        $     &  $8.24 \pm 0.21       $    \\
$\te$ (days)            &   $44.59 \pm 1.03      $    &   $44.35 \pm 0.93       $     &  $45.38 \pm 1.09      $    \\
$s$                     &   $0.9501 \pm 0.0016   $    &   $0.9831 \pm 0.0009    $     &  $1.0285 \pm 0.0014   $    \\
$q$ ($10^{-4}$)         &   $2.630 \pm 0.088     $    &   $2.073 \pm 0.078      $     &  $2.507 \pm 0.087     $    \\
$\alpha$ (rad)          &   $0.521 \pm 0.008     $    &   $0.522 \pm 0.008      $     &  $0.521 \pm 0.007     $    \\
$\rho$ ($10^{-3}$)      &   $0.765 \pm 0.046     $    &   $0.692 \pm 0.038      $     &  $0.674 \pm 0.038     $    \\
\hline
\end{tabular}
%\tablefoot{ ${\rm HJD}^\prime = {\rm HJD}- 2450000$.  }
\end{table*}
% --------------------------------------------------------

% Figure 4 ------------------------------------------------------
\begin{figure}[t]
\includegraphics[width=\columnwidth]{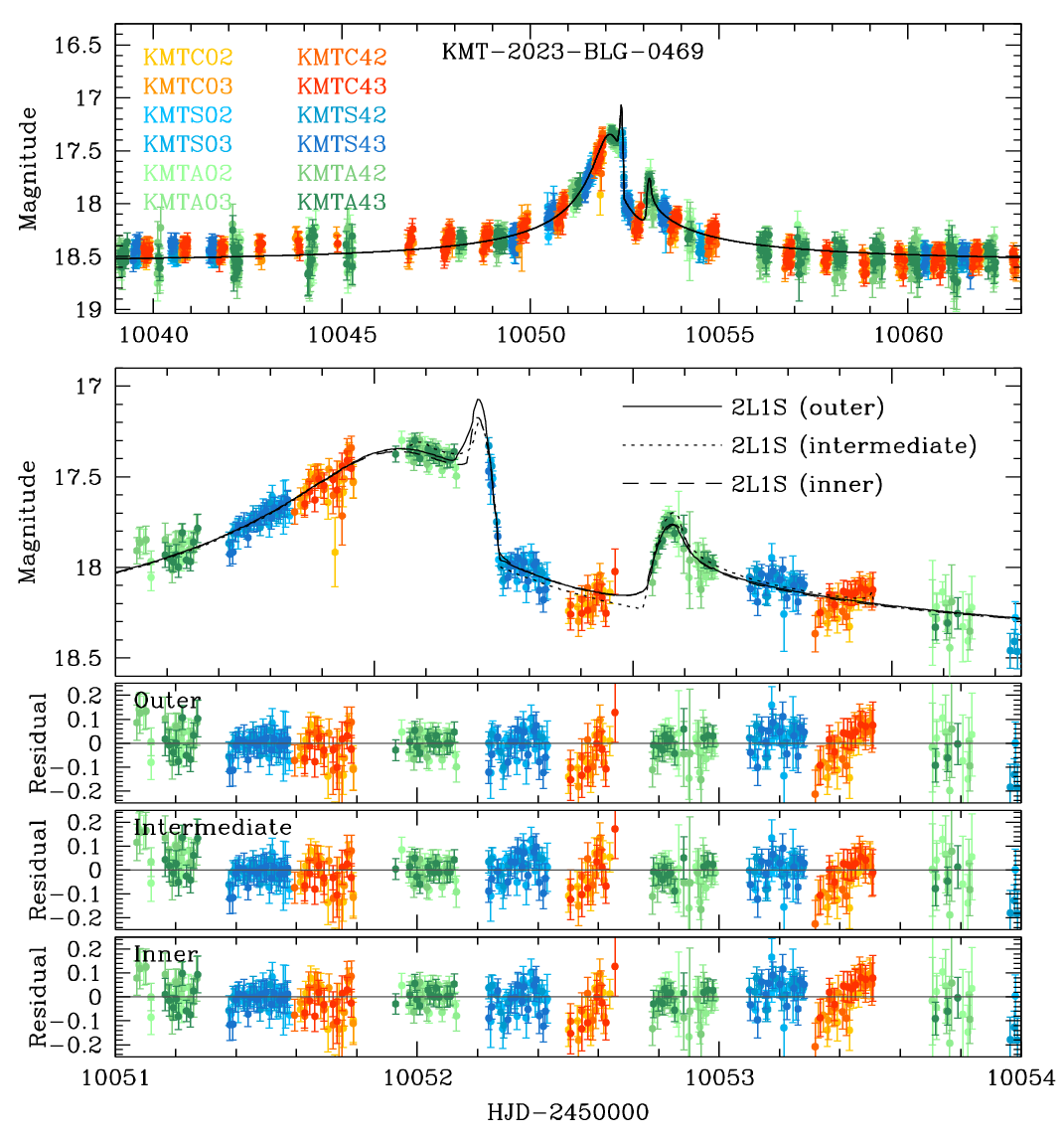}
\caption{
Light curve of KMT-2023-BLG-0469.  The solid, dotted, and dashed curves on the data points 
represent the models of the outer, intermediate, and inner 2L1S solutions, respectively. 
}
\label{fig:four}
\end{figure}
% --------------------------------------------------------------

\subsection{KMT-2023-BLG-0469 }\label{sec:three-two}

The lensing event KMT-2023-BLG-0469 was discovered by the KMTNet group on 2023 April 18 
(${\rm HJD}^\prime \sim 10052$), near which the event reached its maximum magnification of 
$A_{\rm max}\sim 120$.  The source of the event lies in a small (0.8 deg$^2$) region that is 
covered by the four KMTNet prime fields BLG02, BLG03, BLG42, and BLG43, and thus the event was 
observed very densely with a combined cadence of 0.125~hr.  See the layout of the KMTNet fields 
presented in  Figure~12 of \citet{Kim2018}.  As we discuss in Sect.~\ref{sec:four}, the source 
is very faint and the baseline flux is heavily blended.  As a result, the source at the peak was 
brighter than the baseline by only $\sim 1.1$~mag despite the high magnification of the event.

% Figure 5 ------------------------------------------------------
\begin{figure}[t]
\includegraphics[width=\columnwidth]{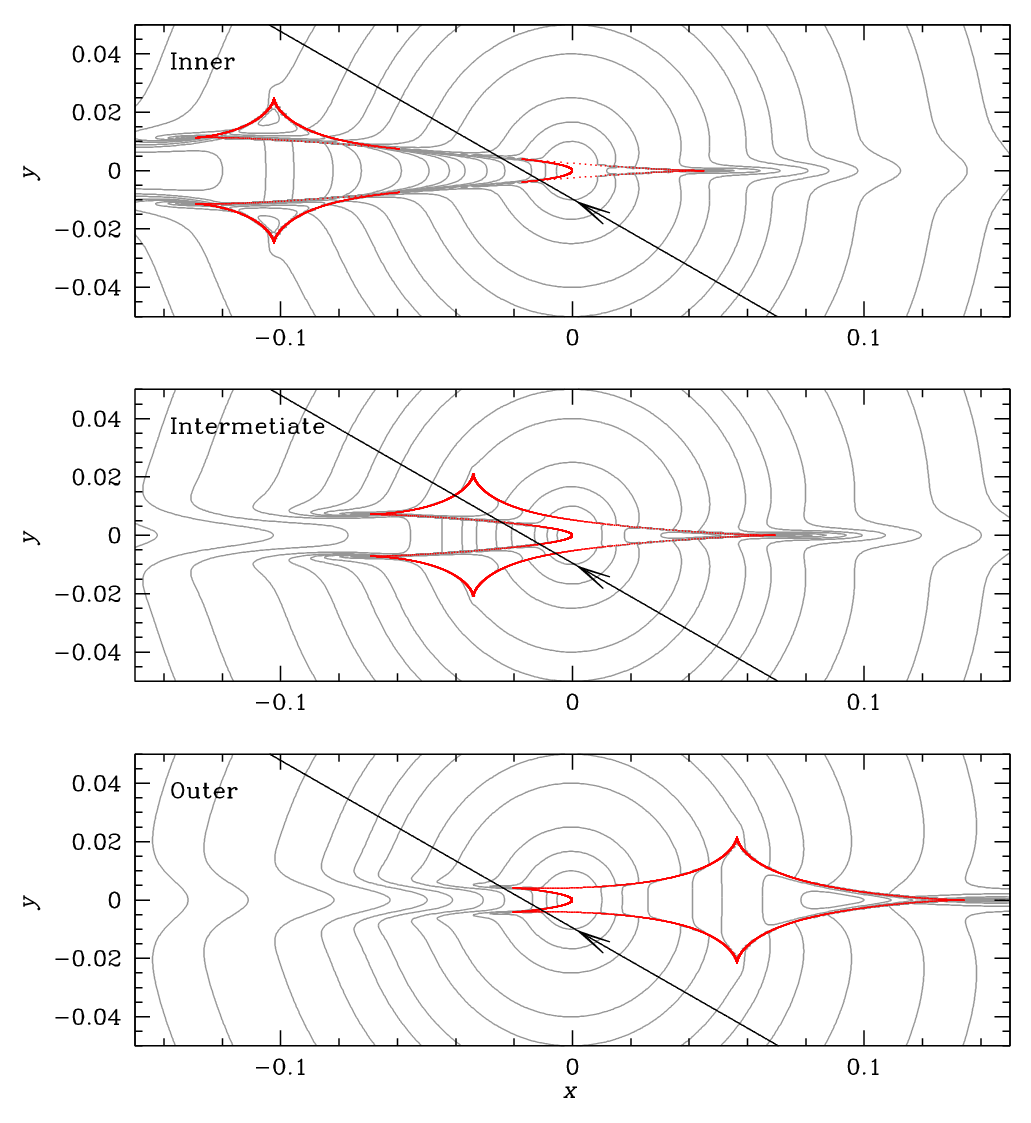}
\caption{
Lens-system configurations of the inner (top panel), intermediate (middle panel), and
outer (bottom panel) solutions for the lensing
event KMT-2023-BLG-0469. Notations are same as those in Fig.~\ref{fig:three}.
}
\label{fig:five}
\end{figure}
% --------------------------------------------------------------

The lensing light curve of KMT-2023-BLG-0469 is shown in Figure~\ref{fig:four}, in which the 
top panel shows the whole view, and the second panel shows the zoom-in view of the anomaly 
region appearing near the peak.  Similar to the case of MOA-2022-BLG-563, the anomaly is 
characterized by a dip feature and the positive peaks appearing before and after the dip. The 
anomaly is centered at $t_{\rm anom}\sim 10052.8$ in HJD$^\prime$, and the time gap between the 
two peaks is $\sim 1.2$~days. The second half of the first peak, which was covered by the KMTS 
data sets, exhibits a sharply declining feature, indicating that the peak was produced by the 
caustic crossing of the source. On the other hand, the second peak exhibits a smoothly varying 
feature, suggesting that the peak was produced by the cusp approach of the source.

From the analysis of the lensing light curve, we find that KMT-2023-BLG-0469 share common
properties to those of the event MOA-2022-BLG-563 in the sense that the anomaly was produced 
by a low-mass planetary companion and the solution is subject to the inner-outer degeneracy. 
The planetary parameters of the inner and outer solutions are $(s, q)_{\rm in}\sim  (0.95, 
2.6\times 10^{-4})$ and $(s, q)_{\rm out}\sim  (1.03, 2.5\times 10^{-4})$, respectively. 
Regardless of the solutions, the estimated planet-to-host mass ratio is much smaller than the 
Jupiter/Sun mass ratio of $\sim 10^{-3}$, and this suggests that the planet is less massive than 
Jupiter. We list the full lensing parameters of the two solutions in Table~\ref{table:three}. 
With the lensing parameters $(t_0, u_0, \te, t_{\rm anom}) \sim  (10052.07, 8.42\times 10^{-3}, 
44.6, 10052.8)$, we find $s^\dagger= 0.991$, which matches well the geometric mean $(s_{\rm in}
\times s_{\rm out})^{1/2}\sim 0.989$, and this indicates that the degeneracy between the two 
solutions is caused by the inner-outer degeneracy.  We mark the position of $s^\dagger$ on the 
$\Delta\chi^2$ map presented in the middle panel of Figure~\ref{fig:two}.

Besides the two solutions resulting from the inner-outer degeneracy, we found an additional 
degenerate solution. The $\Delta\chi^2$ map presented in the middle panel of Figure~\ref{fig:two} 
shows the three local solutions. The additional solution has binary parameters $(s, q)_{\rm int}
\sim (0.98, 2.1\times 10^{-4})$. We designate this solution as "intermediate" solution because the 
planet separation of the solution approximately corresponds to the mean of the inner and outer 
solutions. We list the full lensing parameters of the intermediate solution in Table~\ref{table:three}. 
Among the three degenerate solutions, the outer solution is preferred over the inner and intermediate 
solutions by $\Delta\chi^2 = 12.3$ and 24.3, respectively. In Figure~\ref{fig:four}, we plot the model 
curves and residuals of the outer (solid curve), intermediate (dotted curve), and inner (dashed curve) 
solutions.

In Figure~\ref{fig:five}, we present the lens-system configurations corresponding to the three 
degenerate solutions. The configurations of the pair of the inner and outer solutions have similar 
characteristic to those of the event MOA-2022-BLG-563: the source passed through  the inner region 
between the central and planetary caustics for the inner solution, and the source passed through the 
region outside the caustic for the outer solution.  For the intermediate solution, on the other hand, 
the source passed through the two caustic prongs that extend from the two back-end cusps of  a resonant 
caustic.  According to this solution, the source crossed the caustic four times during the anomaly.  
The first pair of crossings occurred when the source passed through the lower prong, and the other 
pair of crossings occurred when the source passed through the upper prong.  The individual caustic 
crossings produced caustic spikes, but the one produced when the source entered the lower caustic 
prong and the one generated when the source exited the upper caustic prong do not exhibit obvious 
features due to the combination of the weakness of the caustic folds and finite-source effects.  
From the resolved caustic-crossing part of the anomaly, the normalized source radius, $\rho=(0.674 
\pm 0.038)\times 10^{-3}$, is precisely measured.

% Figure 6 ------------------------------------------------------
\begin{figure}[t]
\includegraphics[width=\columnwidth]{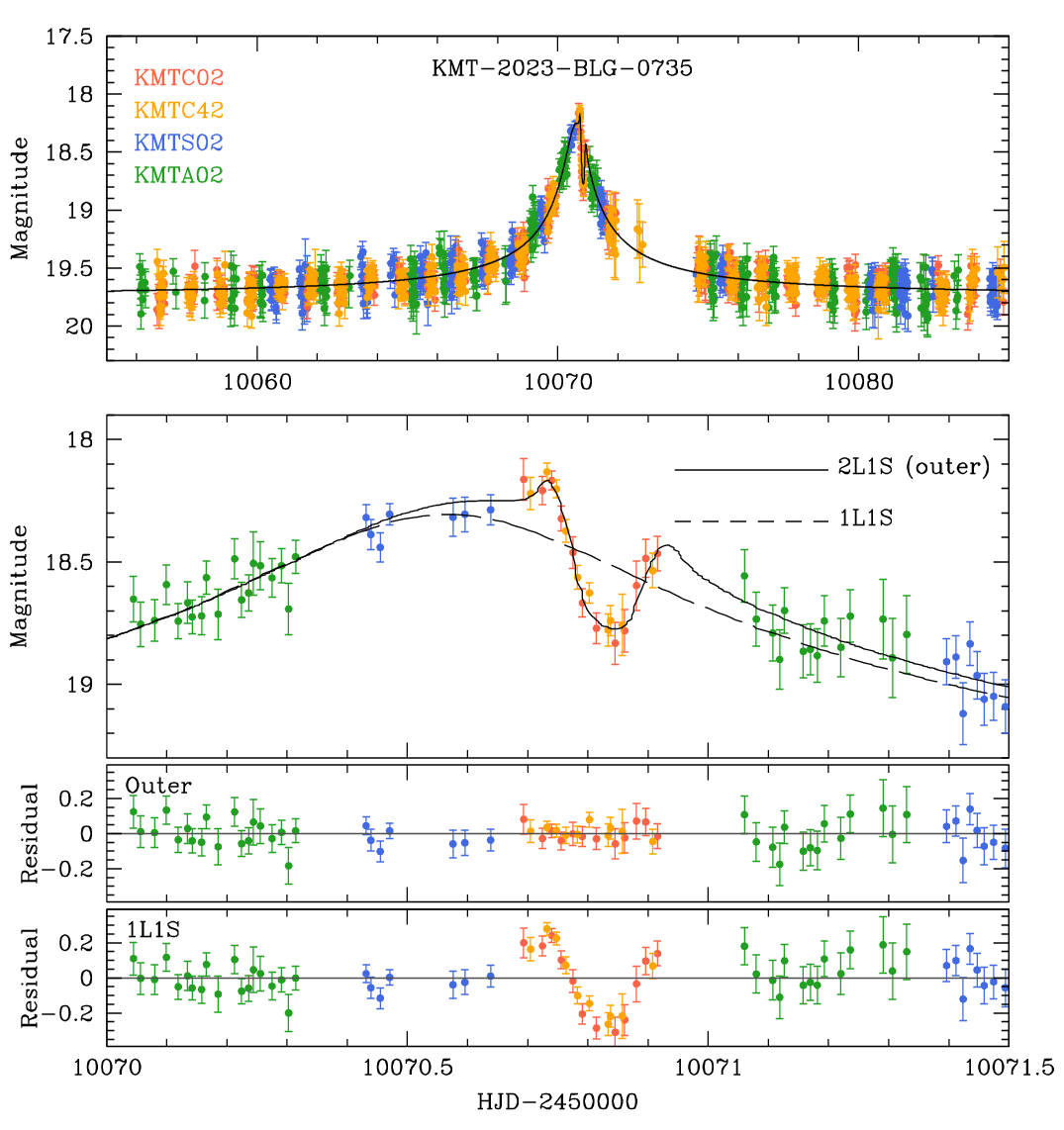}
\caption{
Light curve of KMT-2023-BLG-0735.
}
\label{fig:six}
\end{figure}
% --------------------------------------------------------------

\subsection{KMT-2023-BLG-0735}\label{sec:three-three}

The KMTNet group found the event KMT-2023-BLG-0735 on 2023 May 08 (${\rm HJD}^\prime \sim 10072$), 
which was about two days after the peak of the lensing light curve. The event reached a moderately 
high magnification of $A_{\rm max}\sim 60$ at the peak. The extinction toward the source, $A_I=5.97$, 
is very high because of the closeness of the source to the Galactic center. The source lies near the 
edge of the KMTNet prime fields BLG02 and BLG42.  The KMTC data were recovered from the images of 
both fields, but the KMTS and KMTA data in the BLG42 field were not recovered.  As a result, the 
observational cadence varies depending on the field: 0.25~hr for the KMTC data set and 0.5~hr for 
the KMTS and KMTA data sets.

We present the lensing light curve of KMT-2023-BLG-0735 in Figure~\ref{fig:six}. The light curve
exhibits similar characteristics to those of the previous two events in various aspects. 
First, an anomaly occurred near the peak of the light curve. Second, the anomaly exhibits a dip 
feature, which is centered at $t_{\rm anom}\sim 10070.84$ in HJD$^\prime$, and two positive peaks 
on both sides of the dip. The time gap between the two peak is $\Delta t\sim 0.6$~day. The anomaly 
was covered mostly by the combination of KMTC02 and KMTC42 data sets

% Table 4 ------------------------------------------------
\begin{table}[t]
\small
%\centering
\caption{Model parameters of KMT-2023-BLG-0735.\label{table:four}}
\begin{tabular*}{\columnwidth}{@{\extracolsep{\fill}}llll}
\hline\hline
\multicolumn{1}{c}{Parameter}    &
\multicolumn{1}{c}{Inner}        &
\multicolumn{1}{c}{Outer}       \\
\hline
$\chi^2$/dof            &   $ 2901.8/2908         $       &   $2901.0/2908          $      \\
$t_0$ (HJD$^\prime$)    &   $ 10070.611 \pm 0.009 $       &   $10070.602 \pm 0.011  $      \\
$u_0$ ($10^{-2}$)       &  -$  1.83432 \pm 0.25   $       &  -$ 1.68 \pm 0.283      $      \\
$\te$ (days)            &   $ 17.93 \pm  2.38     $       &   $19.31 \pm 2.28       $      \\
$s$                     &   $ 0.935 \pm 0.009     $       &   $1.041 \pm 0.009      $      \\
$q$ ($10^{-4}$)         &   $ 2.26 \pm  0.39      $       &   $1.92 \pm 0.46        $      \\
$\alpha$ (rad)          &   $ 5.311 \pm 0.029     $       &   $5.334 \pm 0.034      $      \\
$\rho$ ($10^{-3}$)      &   $ 1.64 \pm  0.54      $       &   $1.63 \pm 0.49        $      \\
\hline
\end{tabular*}
\end{table}
% --------------------------------------------------------

% Figure 7 ------------------------------------------------------
\begin{figure}[t]
\includegraphics[width=\columnwidth]{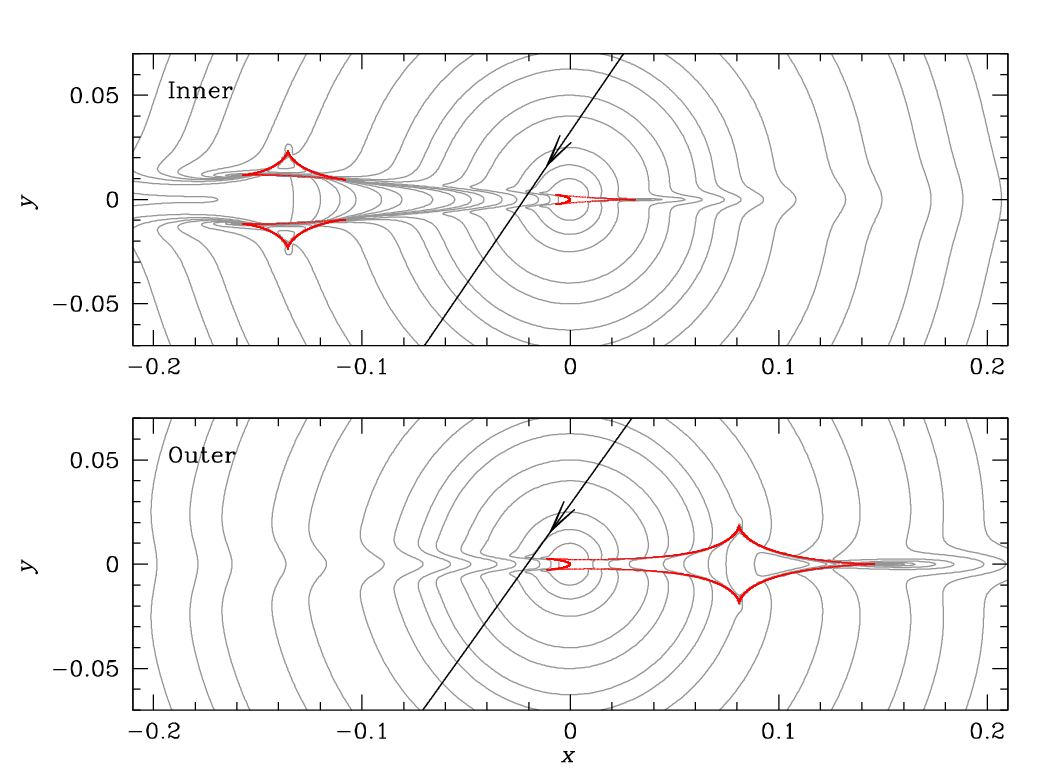}
\caption{
Lens-system configurations of the inner (upper panel) and outer (lower panel) solutions for
the lensing event KMT-2023-BLG-0735. Notations are same as those in Fig.~\ref{fig:three}. 
}
\label{fig:seven}
\end{figure}
% --------------------------------------------------------------

% Figure 8 ------------------------------------------------------
\begin{figure*}[t]
\centering
\includegraphics[width=12.5cm]{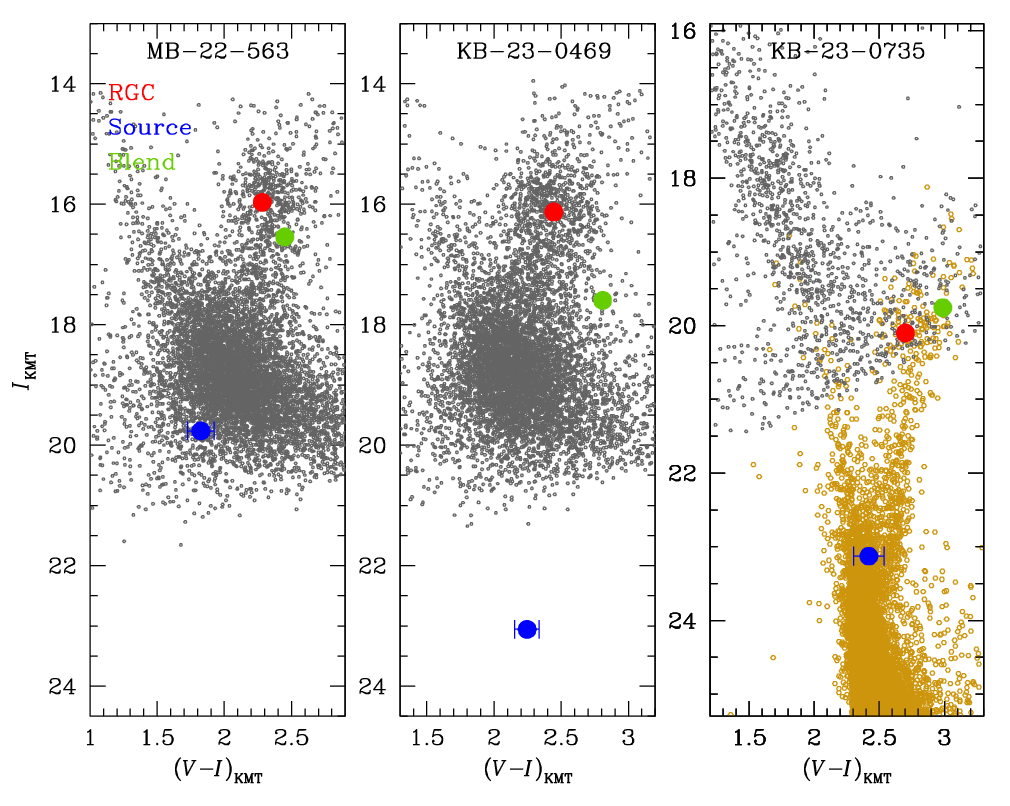}
%\begin{figure}[t]
%\includegraphics[width=\columnwidth]{f8.eps}
\caption{
Locations of the source, blend, and RGC centroid in the instrumental color-magnitude diagrams 
(CMDs) of the lensing events MOA-2022-BLG-563, KMT-2023-BLG-0469, and KMT-2023-BLG-0735.
For KMT-2023-BLG-0735, the CMD is constructed by combining the KMTC data (gray dots) and HST 
(brown dots) data. 
}
\label{fig:eight}
%\end{figure}
\end{figure*}
% --------------------------------------------------------------

% Table 5 ------------------------------------------------
\begin{table*}[t]
\small
%\centering
\caption{Source parameters.  \label{table:five}}
\begin{tabular}{lllll}
%\begin{tabular}{\columnwidth}{@{\extracolsep{\fill}}lllcc}
\hline\hline
\multicolumn{1}{c}{Quantity}                   &
\multicolumn{1}{c}{MOA-2022-BLG-563}           &
\multicolumn{1}{c}{KMT-2023-BLG-0469}          &
\multicolumn{1}{l}{KMT-2023-BLG-0735}          \\
\hline
 $(V-I, I)               $ & $(1.825 \pm 0.100, 19.764 \pm 0.068) $ &  $(2.246 \pm 0.091, 23.056 \pm 0.011)  $ &   $(2.419 \pm 0.117, 23.127 \pm 0.036)$ \\
 $(V-I, I)_{\rm RGC}     $ & $(2.282, 15.967)                     $ &  $(2.446, 16.131)                      $ &   $(2.70, 20.10)                      $ \\
 $(V-I, I)_{{\rm RGC},0} $ & $(1.060, 14.357)                     $ &  $(1.060, 14.384)                      $ &   $(1.060, 14.429)                    $ \\
 $(V-I, I)_0             $ & $(0.604 \pm 0.108, 18.154 \pm 0.071) $ &  $(0.860 \pm 0.100, 21.308 \pm 0.023)  $ &   $(0.779 \pm 0.117, 17.455 \pm 0.036)$ \\
 Spectral type             & F8V                                    &  K0V                                     &   G8 turnoff                            \\
\hline
\end{tabular}
%\tablefoot{ ${\rm HJD}^\prime = {\rm HJD}- 2450000$.  }
\end{table*}
% --------------------------------------------------------

% Table 6 ------------------------------------------------
\begin{table*}[t]
\small
%\centering
\caption{Angular source radius, Einstein radius, and relative lens-source proper motion.  \label{table:six}}
\begin{tabular}{lllll}
%\begin{tabular}{\columnwidth}{@{\extracolsep{\fill}}lllcc}
\hline\hline
\multicolumn{1}{c}{Quantity}                      &
\multicolumn{1}{c}{MOA-2022-BLG-563}                    &
\multicolumn{1}{c}{KMT-2023-BLG-0469}             &
\multicolumn{1}{l}{KMT-2023-BLG-0735}                     \\
\hline
 $\theta_*$ ($\mu$as) & $0.656 \pm  0.084 $ &  $0.204 \pm 0.024 $ &  $1.135 \pm 0.155 $  \\
 $\thetae$ (mas)      & --                  &  $0.302 \pm 0.040 $ &  $0.697 \pm 0.225 $  \\
 $\mu$ (mas/yr)       & --                  &  $2.43 \pm 0.33   $ &  $13.17 \pm 4.54  $  \\
\hline
\end{tabular}
%\tablefoot{ ${\rm HJD}^\prime = {\rm HJD}- 2450000$.  }
\end{table*}
% --------------------------------------------------------

As expected from the similar pattern of the anomaly to those of MOA-2022-BLG-563 and KMT-2023-BLG-0469, 
we find a pair of inner and outer planetary solutions describing the observed anomaly. The planetary 
parameters are $(s, q)_{\rm in}\sim (0.95, 2.6\times 10^{-4})$ for the inner solution and $(s, 
q)_{\rm out}\sim (1.03, 2.5\times 10^{-4})$ for the outer solution, and the estimated mass ratio 
indicates that the planet is likely to be substantially less massive than Jupiter of the Solar 
System.  The positions of the two degenerate solutions in the $\log s$--$\log q$ parameter plane 
are shown on the $\Delta\chi^2$ map presented in the bottom panel of Figure~\ref{fig:two}, and the 
full lensing parameters of both solutions are listed in Table~\ref{table:four}. The two solutions 
are very degenerate, and the outer solution is preferred over the inner solution only by 
$\Delta\chi^2= 0.8$. From the lensing parameters $(t_0, u_0, \te, t_{\rm anom}) \sim  (10070.61, 
1.75\times 10^{-2}, 18.5, 10070.84)$, we find $s^\dagger \sim  0.98$ compared to the geometric mean 
$(s_{\rm in}\times s_{\rm out})^{1/2}\sim 0.97$, indicating that the degeneracy between the solutions 
is caused by the inner-outer degeneracy.  The model curve of the outer solution and the residual are 
shown in Figure~\ref{fig:six}.

Figure~\ref{fig:seven} shows the lens-system configurations corresponding to the inner and outer 
solutions of KMT-2023-BLG-0735. According to the configurations, the anomaly was produced by the 
passage of the source through the negative deviation region on the back-end side of the caustic. 
The configurations are very similar to those of the MOA-2022-BLG-563 in the senses that the ambiguity 
between the two solutions originates from the inner-outer degeneracy, and the central and planetary 
caustics of the inner solution are well separated, while the caustics of the outer solution form a 
single resonant caustic.  Although the source did not cross the caustic, the normalized source radius, 
$\rho=(1.64\pm 0.53)\times 10^{-3}$, is measured because of the large magnification gradient of the 
anomaly region extending from the strong cusp of the caustic.

\section{Source star and Einstein radius}\label{sec:four}

In this section, we characterize the source stars of the events and estimate the angular Einstein 
radii. The Einstein radius is estimated from the relation
\begin{equation}
\thetae = {\theta_*\over \rho},
\label{eq2}
\end{equation}
where the normalized source radius $\rho$ is measured from the light curve modeling, and the angular 
source radius $\theta_*$ is deduced from the type of the source determined from its extinction and
reddening-corrected (dereddened) color and magnitude. Although the angular Einstein radius cannot be 
estimated for MOA-2022-BLG-563 because the normalized source radius could not not measured, we specify 
the source star for the complete characterization of the event.

For each event, the dereddened color and magnitude of the source, $(V-I, I)_0$, were estimated using 
the \citet{Yoo2004} method. In this method, the instrumental color and magnitude, $(V-I, I)$, of the 
source are measured, and then they are calibrated using the centroid of red giant clump (RGC), for 
which its dereddened values, $(V-I, I)_{{\rm RGC},0}$, are known, as a reference, that is,
\begin{equation}
(V-I, I)_0 
= (V-I, I)_{{\rm RGC},0} + \Delta(V-I, I).
\label{eq3}
\end{equation}
Here $\Delta(V-I, I)$ represents the offsets in color and magnitude between the source and RGC centroid. 
The instrumental $V$ and $I$-band magnitudes of the source were measured by regressing the light curves 
of the individual passbands constructed using the pyDIA photometry code \citep{Albrow2017} with respect 
to the model.

For KMT-2023-BLG-0735, the photometry quality of the $V$-band data is not good because of the very 
high extinction toward the field, and this makes it difficult to directly measure the source color.  
In this case, we adopted the median color of stars lying within the range of the measured $I$-band 
magnitude on the main-sequence or giant branches of the combined CMD, which is constructed by 
aligning the CMD constructed using the KMTC data and the CMD of bulge stars lying toward the Baade's 
window constructed from the observations using the Hubble Space Telescope (HST) by \citet{Holtzman1998}.  
For the alignment of the KMTC and HST CMDs, we used $I$-band magnitudes the RGC centroids in the 
individual CMDs as registration marks.  The RGC centroid was chosen as the median position of stars 
in the clump of red giants.  The position of the RGC centroid in the $(I-V)$--$I$ KMTC CMD was somewhat 
uncertain, and thus we determined the $I$-band magnitude of the RGC centroid from the $(I-K)$--$I$ CMD, 
which clearly shows the RGC centroid.  The $(I-K)$--$I$ CMD was constructed by matching stars in the 
KMTC image and those in the catalog of VISTA Variables in the Via Lactea (VVV) survey \citep{Minniti2010}.

In Figure~\ref{fig:eight}, we mark the positions of the source stars (blue dots) and the RGC 
centroids (red dots) of the individual events in the instrumental CMDs of stars lying in the 
vicinity of the source stars.  Also marked in the CMDs are the positions of the blended objects 
(green dots). We found that the source flux is heavily blended by the flux from a nearby giant 
star in all cases of the events.

With the measured instrumental colors and magnitudes together with the de-reddened values of the RGC 
centroid adopted respectively from \citet{Bensby2013} and \citet{Nataf2013}, we then estimated the 
calibrated source colors and magnitudes using the relation in Eq.~(\ref{eq3}). In Table~\ref{table:five}, 
we list the values of $(V-I, I)$, $(V-I, I)_{\rm RGC}$, $(V-I, I)_{{\rm RGC},0}$, and $(V-I, I)_0$ for 
the individual events. According to the estimated color and magnitude, the source star is a late F-type 
main-sequence star for MOA-2022-BLG-563, an early K-type main-sequence star for KMT-2023-BLG-0469, and 
a late G-type turn-off star for KMT-2023-BLG-0735.

The angular source radius is derived from the measured $(V-I, I)_{0}$ by first converting the 
$V-I$ color into $V-K$ color using the color-color relation of \citet{Bessell1988}, and then 
deducing $\theta_*$ from the $(V-K, I)$--$\theta_*$ relation of \citet{Kervella2004}.  With the 
estimated the source radius, the angular Einstein radius is then estimated using the relation in 
Eq.~(\ref{eq2}), and the relative lens-source proper motion is estimated as 
\begin{equation}
\mu = {\thetae  \over \te}.
\label{eq4}
\end{equation}
In Table~\ref{table:six}, we list the estimated values of $\theta_*$, $\thetae$, and $\mu$ for the 
individual lensing events.

% Table 7 ------------------------------------------------
\begin{table*}[t]
\small
%\centering
\caption{Physical lens parameters. \label{table:seven}}
\begin{tabular}{lcccc}
%\begin{tabular}{\columnwidth}{@{\extracolsep{\fill}}lllcc}
\hline\hline
\multicolumn{1}{c}{Parameter}                 &
\multicolumn{1}{c}{MOA-2022-BLG-563}          &
\multicolumn{1}{c}{KMT-2023-BLG-0469}         &
\multicolumn{1}{l}{KMT-2023-BLG-0735}         \\
\hline
 $M_{\rm h}$ ($M_\odot$)      &   $0.48^{+0.36}_{-0.30}$     &  $0.47^{+0.35}_{-0.26}   $   &   $0.61^{+0.34}_{-0.34}   $    \\  [0.6ex]
 $M_{\rm p}$ ($M_{\rm J}$)    &   $0.40^{+0.31}_{-0.25}$     &  $0.124^{+0.092}_{-0.067}$   &   $0.122^{+0.068}_{-0.068}$    \\  [0.6ex]
 $\dl$ (kpc)                  &   $6.53^{+1.12}_{-1.57}$     &  $7.07^{+1.03}_{-1.19}   $   &   $6.39^{+1.21}_{-1.60}   $    \\  [0.6ex]
 $a_{\perp,{\rm in}}$ (AU)    &   $2.15^{+0.37}_{-0.52}$     &  $2.32^{+0.33}_{-0.38}   $   &   $2.66^{+0.50}_{-0.66}   $    \\  [0.6ex]
 $a_{\perp,{\rm out}}$ (AU)   &   $2.40^{+0.41}_{-0.58}$     &  $2.42^{+0.35}_{-0.41}   $   &   $2.96^{+0.56}_{-0.74}   $    \\  [0.6ex]
 $a_{\perp,{\rm int}}$ (AU)   &   --                         &  $2.31^{+0.34}_{-0.39}   $   &   --                           \\  [0.6ex]
 $a_{\rm snow}$ (AU)          &   1.3                        &  1.3                         &   1.67                         \\
 $P_{\rm disk}$               &   36\%                       &  27\%                        &   42\%                         \\
 $P_{\rm bulge}$              &   64\%                       &  73\%                        &   58\%                         \\
\hline
\end{tabular}
%\tablefoot{ ${\rm HJD}^\prime = {\rm HJD}- 2450000$.  }
\end{table*}
% --------------------------------------------------------

% Figure 9 ------------------------------------------------------
\begin{figure}[t]
\includegraphics[width=\columnwidth]{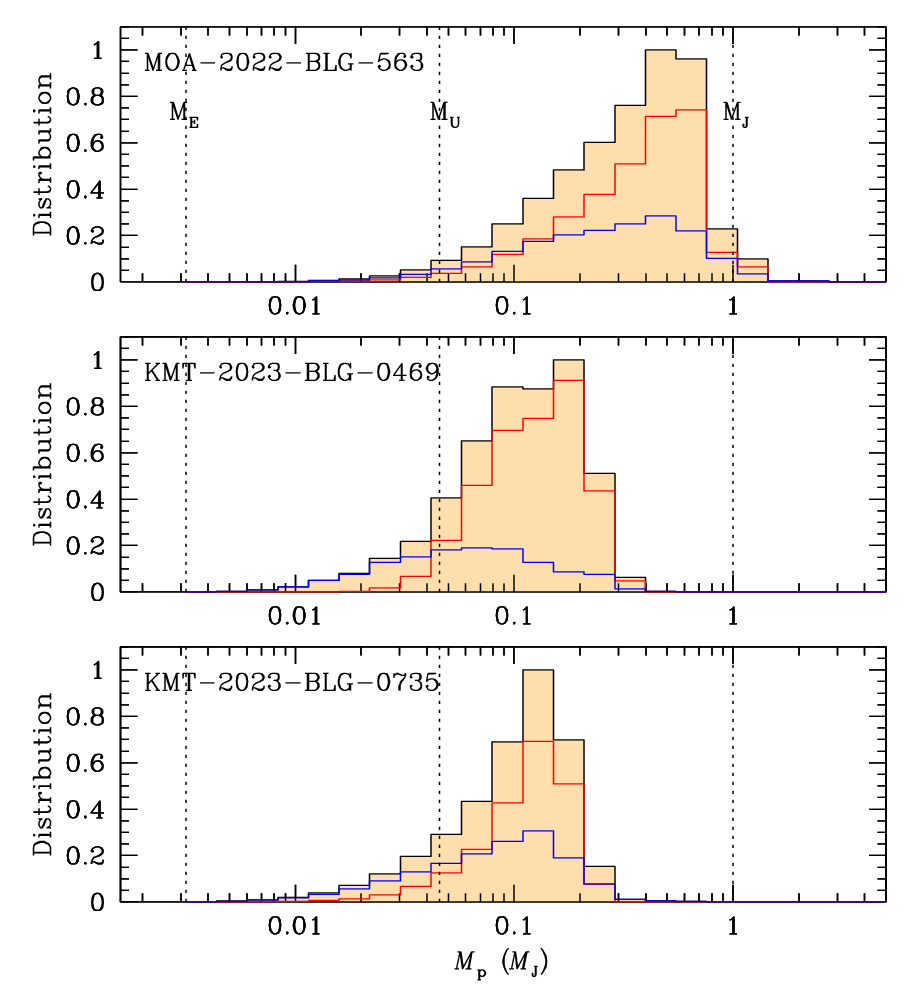}
\caption{
Bayesian posteriors of the planet masses. The three dotted vertical lines represent the masses 
of Earth ($M_{\rm E}$), Uranus ($M_{\rm U}$), and Jupiter ($M_{\rm J}$) of the Solar System. 
The curves drawn in red and blue indicate the contributions  by the disk and bulge lens 
	populations, respectively, and the black curve represents the sum of the contributions 
	by the two lens populations.
}
\label{fig:nine}
\end{figure}
% --------------------------------------------------------------

\section{Physical lens parameters}\label{sec:five}

The physical parameters of the individual planetary systems were estimated from Bayesian analyses,
which were conducted using the constraints given by the measured lensing observables. In the first
step of the analysis, we produced a large number of artificial lensing events from a Monte Carlo
simulation.  For each simulated event, we assigned the mass $M$ of the lens, which was derived from 
a model mass function, and the distances to the lens $\dl$ and source $\ds$, and their relative 
proper motion $\mu$, which were derived from a Galaxy model.  In the simulation, we adopted the 
\citet{Jung2018} model for the mass function, and the \citet{Jung2021} model for the Galaxy model. 
In the second step, we calculated the lensing observables of the event time scale and Einstein 
radius corresponding to the physical parameters $(M, \dl, \ds, \mu)$ of each simulated event from 
the relations 
\begin{equation}
\te={\thetae \over \mu};\qquad
\thetae = (\kappa M \pi_{\rm rel})^{1/2},
\label{eq5}
\end{equation}
where $\kappa =4G/(c^2{\rm AU})\simeq 8.14~{\rm mas}/M_\odot$, and $\pi_{\rm rel} = \pi_{\rm L}-
\pi_{\rm S}= {\rm AU}(1/\dl - 1/\ds)$ denotes the relative lens-source parallax. For all three
events, the extra lensing observable of the microlens parallax $\pie$ could not be constrained, 
because either the event time scale was not long enough or the photometric precision of data was 
not high enough to detect the subtle deviations induced by the higher-order effects. In the third 
step, we constructed the posteriors of the lens mass and distance by imposing a weight to each 
event of
\begin{equation}
w_i = \exp\left(-{\chi_i^2 \over 2} \right);\qquad
\chi_i^2 = 
{(t_{{\rm E},i}-\te)^2 \over \sigma^2(\te)} + 
{(\theta_{{\rm E},i}-\thetae)^2 \over \sigma^2(\thetae)},
\label{eq6}
\end{equation}
where $(\te, \thetae)$ denote the measured values of the lensing observables, and $[\sigma(\te), 
\sigma(\thetae)]$ represent their uncertainties. Besides the lensing observables $\te$ and $\thetae$, 
the blended flux can often provide a constraint on the lens system from the fact that the flux from 
the lens contributes to the blended light, and thus it should be less than the measured blended flux. 
We found that the blending constraint has little effects on the lens parameters, because the blended 
flux was high in all cases of the events.

% Figure 10 ------------------------------------------------------
\begin{figure}[t]
\includegraphics[width=\columnwidth]{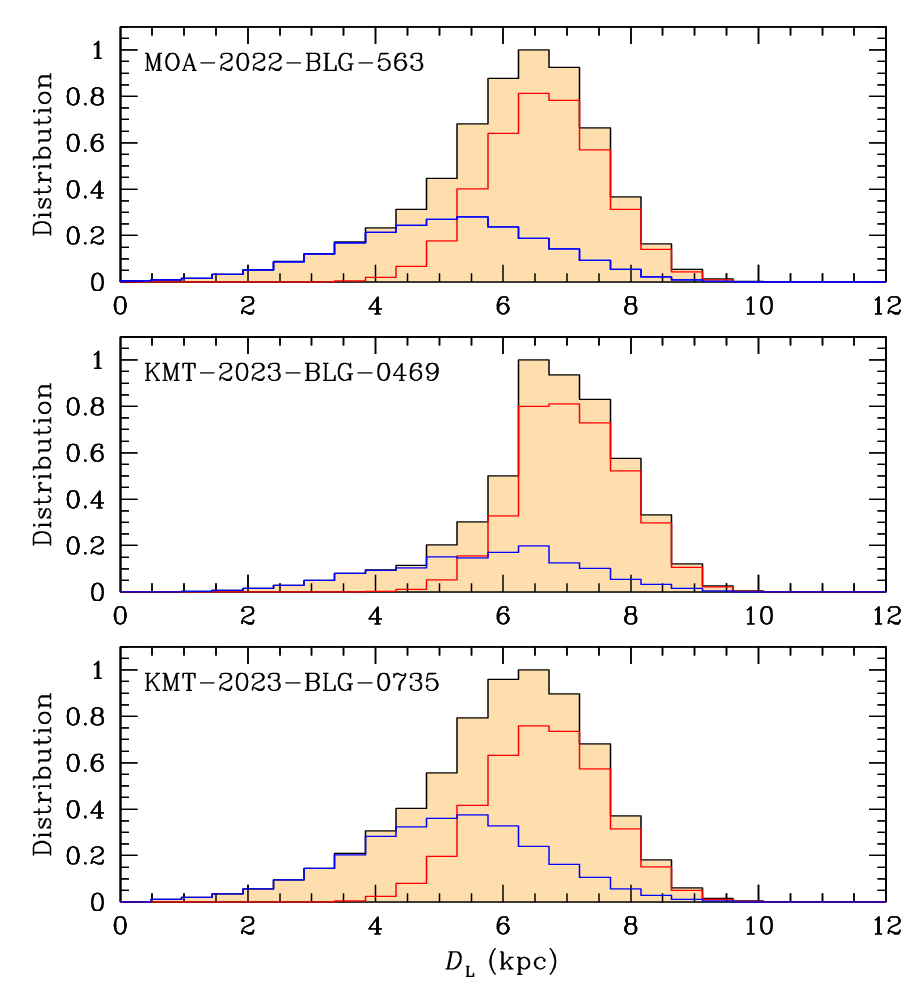}
\caption{
Bayesian posteriors of the distances to the planetary systems.  Notations are the same as those 
in Fig.~\ref{fig:nine}.
}
\label{fig:ten}
\end{figure}
% --------------------------------------------------------------

We present the posterior distributions of the planet mass and distance to the planetary lens
system for the individual events in Figures~\ref{fig:nine} and \ref{fig:ten}, respectively. 
In each panel, the distributions drawn in blue and red represent the contributions by the disk 
and bulge lens populations, respectively, and the distribution drawn in black represents the sum 
of contributions by the two lens populations. In the panels of the planet-mass posteriors, we draw 
three dotted vertical lines corresponding to the masses of Jupiter ($M_{\rm J}$), Uranus 
($M_{\rm U}$), and Earth ($M_{\rm E}$) of the Solar System.

In Table~\ref{table:seven}, we list the physical parameters the host mass, $M_{\rm h}$, planet 
mass, $M_{\rm p}$, distance to the planetary system, $\dl$, and the projected planet-host 
separation, $a_\perp$. For the planetary separation, we present the values corresponding to the 
inner ($a_{\perp,{\rm in}}$) and outer ($a_{\perp,{\rm out}}$) solutions, and additionally present 
the value corresponding to the intermediate solution ($a_{\perp,{\rm int}}$) for KMT-2023-BLG-0469. 
For each lens parameter, the median of the Bayesian posterior distribution is chosen as the 
representative value, and the 16\% and 84\% of the posterior distribution are set as the upper and 
lower limits, respectively.  Also presented are the snow line distances estimated as $a_{\rm snow}
\sim 2.7 {\rm AU}(M/M_\odot)$ \citep{Kennedy2008}, and the probabilities for the lenses to be in 
the disk, $P_{\rm disk}$, and bulge, $P_{\rm bulge}$.

It is found that all three planets have masses lying between those of Jupiter and Uranus.  For 
all planetary systems, the hosts of planets are main-sequence stars that are less massive than 
the Sun, and the planets lie well beyond the snow lines of the planetary systems. For the planet 
MOA-2022-BLG-563Lb, the estimated mass is 1.3 times of the mass of Saturn, and thus it is likely 
to be a giant planet. On the other hand, the masses of the planets KMT-2023-BLG-0469Lb and 
KMT-2023-BLG-0735Lb are approximately 2.7 times of the mass of Uranus, and thus they are more 
likely to be ice giants.  In all cases, the planetary systems are more likely to be in the bulge 
with probabilities $P_{\rm bulge} = 64\%$, 73\%, and 58\% for MOA-2022-BLG-563, KMT-2023-BLG-0469, 
and KMT-2023-BLG-0735, respectively.

\section{Summary and conclusion}\label{sec:six}

We conducted analyses of the three microlensing events MOA-2022-BLG-563, KMT-2023-BLG-0469, and 
KMT-2023-BLG-0735, for which the lensing light curves exhibit short-term anomalies with dip features 
appearing near the peaks of the lensing light curves. From the detailed analyses of the anomalies, 
we found that the anomalies were produced by planets, for which the mass ratios between the planet 
and host are on the order of $10^{-4}$.  All planetary systems share common properties that planets 
have masses lying between those of Jupiter and Uranus, the hosts of planets are main-sequence stars 
that are less massive than the Sun, and the planets lie beyond the snow lines of the planetary 
systems.  We found that interpreting the anomalies was subject to a common inner-outer degeneracy, 
which causes ambiguity in estimating the projected planet-host separation, and identified an extra 
local solution resulting from an accidental degeneracy in the case of KMT-2023-BLG-0469.

Conducting morphological studies to classify lensing events based on similar anomalies and 
exploring the specific origins of each anomaly class is crucial. This not only aids in promptly 
diagnosing anomalies before in-depth analysis but also facilitates an accurate characterization 
of lens systems for future events sharing similar anomaly structures. The anomalies observed in 
the analyzed events share a distinctive trait, characterized by a brief dip surrounded by subtle 
bumps near the peak of the light curve. This particular feature strongly suggests a planetary 
origin for the anomaly.

% --------------------------------------------------------------
\begin{acknowledgements}
Work by C.H. was supported by the grants of National Research Foundation of Korea 
(2019R1A2C2085965).
% KMTNet
This research has made use of the KMTNet system operated by the Korea Astronomy and Space Science 
Institute (KASI) at three host sites of CTIO in Chile, SAAO in South Africa, and SSO in Australia. 
Data transfer from the host site to KASI was supported by the Korea Research Environment Open NETwork 
(KREONET).
This research was supported by the Korea Astronomy and Space Science Institute under the R\&D
program (Project No. 2023-1-832-03) supervised by the Ministry of Science and ICT.
% MOA
The MOA project is supported by JSPS KAKENHI Grant Number JP24253004, JP26247023, JP23340064, 
JP15H00781, JP16H06287, JP17H02871 and JP22H00153.
%Yee
J.C.Y., I.G.S., and S.J.C. acknowledge support from NSF Grant No. AST-2108414. 
%Yossi Shvartzvald
Y.S.  acknowledges support from NSF Grant No. 2020740.
% Clement Ranc
C.R. was supported by the Research fellowship of the Alexander von Humboldt Foundation.
\end{acknowledgements}

\end{document}